\documentclass[12pt]{article}
\usepackage{url}
\usepackage{times}
\usepackage{amsmath,amssymb}
\usepackage{amsfonts}

\newcommand{\seq}[1]{\langle #1\rangle}

\newcommand{\set}[1]{\left\{ #1\right\}}

\newcommand{\Def}{:=}

\newcommand{\realrange}[2]{\left[#1, #2\right]}

\newcommand{\unitrange}[2]{\realrange{0}{1}}

\newcommand{\Oh}[1]{\mathcal{O}\!\left( #1\right)}

\newcommand{\Th}[1]{\Theta\!\left( #1\right)}

\newcommand{\llabel}[1]{\label{\labelprefix:#1}}
\newcommand{\labelprefix}{} % later redefined using renewcommand

\newcommand{\discussionsize}{\small}

\newenvironment{code}{\noindent%\sf%
\begin{tabbing}%
\hspace{2em}\=\hspace{2em}\=\hspace{2em}\=\hspace{2em}\=\hspace{2em}\=%
\hspace{2em}\=\hspace{2em}\=\hspace{2em}\=\hspace{2em}\=\hspace{2em}\=%
\kill}{\end{tabbing}}

\newcommand{\labelcommand}{}
\newcommand{\captiontext}{}
\newsavebox{\codeparam}
\newcounter{lineNumber}
\newenvironment{disscodepos}[3]{%
\renewcommand{\labelcommand}{#2}%
\renewcommand{\captiontext}{#3}%
\sbox{\codeparam}{\parbox{\textwidth}{#3}}%
\begin{figure}[#1]\begin{center}\begin{code}\setcounter{lineNumber}{1}}{%
\end{code}\end{center}\caption{\llabel{\labelcommand}\captiontext}\end{figure}}

{\end{disscodepos}}

\newcommand{\While}    {{\bf while\ }}

\newcommand{\Do}       {{\bf do\ }}

\newcommand{\For}      {{\bf for\ }}

\newcommand{\If}       {{\bf if\ }}
\newcommand{\Is}       {:=}

\newcommand{\Then}     {{\bf then\ }}

\newcommand{\RRem}[1]   {\`{\bf --\hspace{0.5mm}--~}{\rm#1}}

\newcommand{\Increment}{\raisebox{.12ex}{\hbox{\tt ++}}}

\newdimen\endofsize\endofsize=0.5em
\def\endofbeweis{~\quad\hglue\hsize minus\hsize
                 \hbox{\vrule height \endofsize width
\endofsize}\par}
\usepackage{tikz}
\usepackage{pgfplots}
\usepackage{microtype}
\newcommand{\pdach}{\overline{p}}

\newcommand{\nopetaedge}[1]{}
\newcommand{\fragelater}[1]{}
\definecolor{mygray}{rgb}{0.5,0.5,0.5}
\definecolor{myred}{rgb}{1,0.5,0.5}
\definecolor{myblue}{rgb}{0.5,0.5,1}
\definecolor{mygreen}{rgb}{0.4,0.8,0.4}

\newcounter{lastnote}

\title{Linear Work Generation of R-MAT Graphs}
\author
{Lorenz H\"ubschle-Schneider, Peter Sanders\\
\\
\normalsize{Karlsruhe Institute of Technology, 76128 Karlsruhe, Germany}\\
\\
\normalsize{\url{[sanders,huebschle]@kit.edu}.}
}

\date{}

\begin{document}
\maketitle
\begin{abstract}
  R-MAT is a simple, widely used recursive model for generating ``complex
  network'' graphs with a power law degree distribution and community
  structure. We make R-MAT even more useful by reducing the required work
  per edge from logarithmic to constant. The algorithm works in an embarrassingly parallel way.
\end{abstract}

\section{Introduction}\label{s:intro}

Algorithms for processing large graphs have attracted a lot of
interest in the last years since many ``big data'' problems can be
described with this abstraction. However, one limitation in developing
these algorithms is in obtaining sufficiently large inputs.  Today, the
largest real world networks, e.g., the Facebook graph, are not available
to most researchers.  When we want to anticipate future
developments, no real world inputs are available at all. Hence,
scalable graph generators are an important surrogate;
e.g., \cite{chakrabarti2004r,San16,FLSSSL18}.  A flagship
model with respect to large graphs is the R-MAT model
\cite{chakrabarti2004r}.  R-MAT is simple, models power law degree
distributions, and also has a community structure that is somewhat similar to
complex real world networks. R-MAT is therefore used in the Graph~500
benchmark \cite{graph500} that is used to evaluate graph algorithms on
the largest massively parallel machines.

The simplest variant of R-MAT defines a directed multi-graph
$G=(V,E)$ with $n=2^k$ nodes and $m$ edges based on 4 parameters $a$,
$b$, $c$, and $d$ with $a+b+c+d=1$. Each edge is drawn independently at
random using the following recursive process: the adjacency matrix is
split into four quadrants.  The edge is placed in the upper left
quadrant with probability $a$, in the upper right quadrant with
probability $b$, in the lower left one with probability $c$ and in the
lower right one with probability $d$. The process repeats this
subdivision $k$ times until a single entry of the adjacency matrix is
determined.
%; see also \frage{Figure~\ref{fig:quadrants}}.
In other words, a for-loop generates one bit of the column and row
indices for the edge in each iteration.  Our main result -- described in
Section~\ref{s:algorithm} -- is to explain how a logarithmic number of
address bits can be generated in each iteration without changing the
underlying process. Section~\ref{s:experiments} reports experiments
which show that the new process is an order of magnitude
faster than the most widely used R-MAT generator \cite{graph500}.
Refer to Section~\ref{s:generalizations} for generalizations to
undirected graphs, more general stochastic Kronecker graphs, bipartite
graphs, simple graphs, smoothed degree distribution, massively
parallel implementation, partitioned graphs, and adaptations to GPUs.

%%%%%%%%%%%%%%%%%%%%%%%%%%%%%%%%%%%%%%%%%%%%%%%%%%%%%%%%%%%%%%%%%%%%%%
\section{Our Algorithm}\label{s:algorithm}

We exploit that one can sample in constant time from any probability distribution
on a finite set. One uses a data structure $A$ that can be built in time proportional to
the number of elements in the set \cite{walker1977alias}.  The basic idea is to precompute $n^{\alpha}$
paths together with their probabilities in the recursive process
described in Section~\ref{s:intro} for some constant $\alpha<1$.
By sampling and concatenating the resulting paths, we generate
$\Th{\log n}$ address bits of adjacency array entries in each iteration.
The simplest such strategy generates all paths of a fixed
length $\ell<0.5\log n$.

We can increase the average number
of generated bits by choosing paths of similar probability.
For example, for a given bound $n$ on the table size,
the following code fragment produces a set of entries $Q$ that maximizes the minimum probability of entries:
%% for a parameter $\pdach$, we can recursively generate paths with probability in the range $[\min\set{a,b,c,d}\pdach,\pdach]$ as follows:

%% \begin{code}
%% \Procedure enumItems$(i,j:\set{0,1}^*,p:\rplus)$\RRem{Initially $(\seq{},\seq{},1)$}\+\\
%%   \Rem{Enumerate items whose row and column address begin with $i$ and $j$, respectively.}\\
%%   \Rem{The items with this property have total probability $p$.}\\
%%   \If $p\leq \pdach$ \Then output item $(i,j,p)$\\
%%   \Else\>\kern0.6em enumItems$(i\cdot0,j\cdot0, ap)$;\ enumItems$(i\cdot 0,j\cdot 1, bp)$\\
%%        \>\kern0.6em enumItems$(i\cdot1,j\cdot0, cp)$;\ enumItems$(i\cdot 1,j\cdot 1, dp)$
%% \end{code}
\begin{code}
%\Function items$(n, a, b, c, d)$\+\\
  $Q\Is\set{(1,\seq{},\seq{})}$\RRem{max-priority queue of possible table entries}\\
  \While $|Q|< n^{\alpha}-4$ \Do\+\\
    $(p,i,j)\Is Q$.deleteMax\RRem{expand maximum probability entry}\\
    $Q.$insert$(ap, i\cdot 0,j\cdot 0)$;\quad $Q.$insert$(bp, i\cdot 0, j\cdot 1)$;\RRem{``$\cdot$'' stands for}\\
    $Q.$insert$(cp, i\cdot 1,j\cdot 0)$;\quad $Q.$insert$(dp, i\cdot 1, j\cdot 1)$\RRem{string concatenation}
%  \Return $Q$
\end{code}
This pseudocode uses strings of bits for clarity where a real
implementation processes machine words in a bit parallel fashion
in constant time.

The resulting sequence of output items $E$ is then preprocessed into a data structure $A$ allowing
constant time sampling. Row and column indices of adjacency array entries can then be generated as follows:
\begin{code}\sf
  $i\Def j\Def\seq{}$\RRem{}fragments of row and column index bit strings\\
  \For $(e\Def 1;\quad $\ $e<B$; $)$\RRem{generate batch of $B$ edges}\+\\
    $(i',j')\Def A.sample$\RRem{get more bits}\\
    $i\Def i\cdot i'$;\quad $j\Def j\cdot j'$\RRem{append them to known bits}\\
    \If $|i|\geq k$ \Then\+\RRem{enough bits for a new edge}\\
      output edge $(i[0..k-1], j[0..k-1])$\\
      $i\Def i[k..]$;\quad  $j\Def j[k..]$\RRem{reuse remaining bits for}\\
      $e\Increment$\RRem{next edge}
\end{code}

We use an information theoretic argument to understand why generating variable rather than fixed length fragments of row/column indices can help. Sampling from $A$
yields $\geq\log(1/\pdach)$ bits of information where $\pdach$ is the maximum probability of an entry in $A$. This table will have size $\approx 1/\pdach$.
Sampling a single
bit of row and column index yields
$H=-a\log a-b\log b-c\log c-d\log d\leq 2$ bits of information -- the entropy of the $4$-digit
alphabet with probabilities $a$, $b$, $c$, $d$.
A table of size $|A|=1/\pdach$ that stores equal length fragments of row/column indices
can store fragments of size $\log(1/\pdach)/2$. This yields $H\log(1/\pdach)/2$ bits of information.
Taking the ratio of these two numbers, we get the possible speedup through using variable length fragments
$\log(1/\pdach)/(H\log(1/\pdach)/2)=2/H\geq 1$.

For example, the Graph 500 benchmark uses $a=0.57$, $b=c=0.19$, and
$d=0.05$. This distribution has entropy $H=1.59$, i.e., the method
with variable length entries for $A$ needs a factor of about 1.26 times less samples from $A$.
Higher gains
are possible for more skewed parameters $a$--$d$.

%%%%%%%%%%%%%%%%%%%%%%%%%%%%%%%%%%%%%%%%%%%%%%%%%%%%%%%%%%%%%%%%%%%%%%
\section{Generalizations}\label{s:generalizations}

Several generalization are quite easy. We can generate an {\bf
  undirected graph} by mirroring edges in the lower left triangular
matrix to the upper right triangular matrix.
The Graph 500 benchmark {\bf scrambles} row and column IDs in order to ``hide'' the structure of the graph. This is a simple constant time postprocessing.
More general {\bf stochastic Kronecker graphs} \cite{LesFal07}
can be generated by using a $k\times k$-matrix rather than a $2\times
2$-matrix for the recursion. We can generate {\bf bipartite graphs} or
{\bf remove duplicate edges} as proposed in \cite{chakrabarti2004r} or
avoid them altogether as described below for generating partitioned
graphs.

Chakrabarti \cite{chakrabarti2004r} proposes to {\bf smooth} the
degree distribution of the graph by perturbing the parameters $a$--$d$ in
each step of the process. Note that this step is often ignored -- for example in
the Graph 500 benchmark. Implementing this exact approach in constant
time per edge seems difficult. However, we can perturb the
probabilities for the entries of $A$ or also use several such
perturbed tables to achieve a similar effect.

\paragraph*{Parallelization.}
Edge generation in our algorithm is embarrassingly parallel and
communi\-cation-free.  Here we note that one could also achieve a
classical goal of parallel algorithm theory to have not only linear
work but also logarithmic time on the critical path (span) for an
unbounded number of processors.
This can be
achieved by generating the entries of $A$ by recursively expanding all entries
with probability $>\pdach$ for an appropriately choosen value of $\pdach$.
This can be done with span $\log_{1/\max\set{a,b,c,d}}(1/\pdach)$ and work linear in $|A|$.
We can then use the parallel algorithm for table
construction from \cite{HubSan19} which has span $\Oh{\log|A|}$ and
linear work.

More practically relevant is that parallel graph algorithms work best
if each PE is responsible for all edges incident to a given number of
nodes.  Sorting the edges accordingly after they were generated would
destroy the communication-free property of our generator.  We can get
a {\bf communication-free} generator for such {\bf partitioned graphs}
using an approach similar to the one used for Erd{\H o}s-R{\'e}nyi
graphs in \cite{FLSSSL18}.
The adjacency matrix is split into a number of quadratic \emph{tiles}.
We use a divide-and-conquer algorithm that
determines the number of edges generated in each of these tiles. Each PE only recurses on those subproblems that
intersect the rows of the adjacency matrix assigned to it. The base
case is a single tile of the adjacency matrix where we use the
algorithm from Section~\ref{s:algorithm} to generate the prescribed
number of entries. Note that this makes duplicate elimination a local
operation so that this feature can also be implemented in a
communication-free way.

Finally, the algorithms are sufficiently simple to be easy to port to
GPU.\@  Perhaps, the single sampling loop from Section~\ref{s:algorithm} would
then be split into separate loops for generating random numbers,
retrieving bit strings from $A$, and chopping the resulting long bit
strings into edges.

%%%%%%%%%%%%%%%%%%%%%%%%%%%%%%%%%%%%%%%%%%%%%%%%%%%%%%%%%%%%%%%%%%%%%%
\section{Experiments}\label{s:experiments}

% RANGE BEGIN plot
% CONNECT sqlite:/global_data/lorenz/wrs/results.db
% SQL CREATE INDEX IF NOT EXISTS rmattype ON rmat (type)
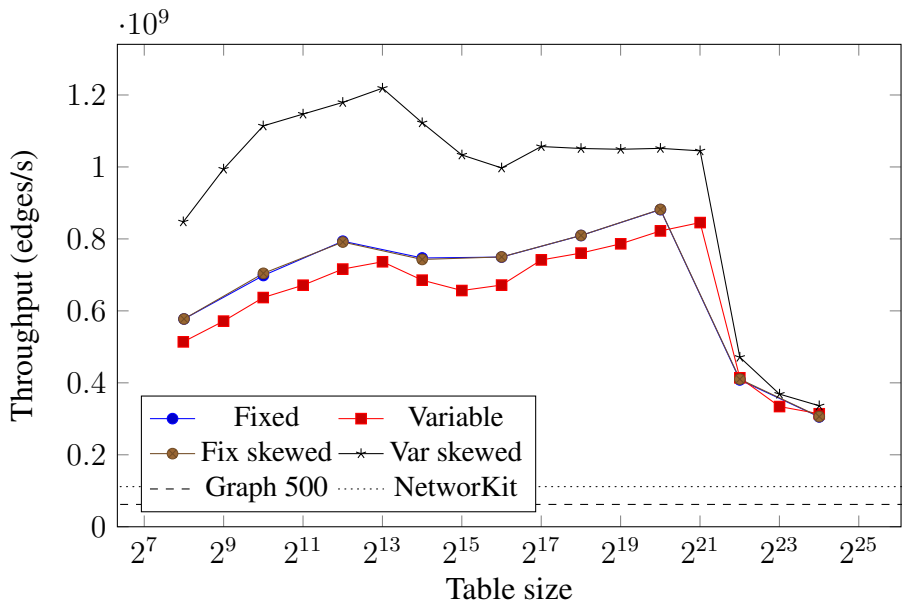
\begin{figure}[b]
\centering
\begin{tikzpicture}
  \begin{semilogxaxis}[
    log basis x=2,
    xlabel={Table size},
    ylabel={Throughput (edges/s)},
    legend pos=south west,
    legend columns=2,
    legend style={font=\small},
    width=12cm, % for now
    height=8cm,
    ymin=0,
    xmin=80,
    xmax=70000000
  ]
  %% MULTIPLOT(method|title) SELECT method as title, paths AS x,
  %% edges / avg(total) * 1000 AS y, MULTIPLOT
  %% FROM rmat WHERE type="sampling" AND method != "graph500" AND method != "networkit"
  %% GROUP BY MULTIPLOT, paths ORDER BY MULTIPLOT, paths
  \addplot coordinates { (256,5.77454e+08) (1024,6.98344e+08) (4096,7.93657e+08) (16384,7.47047e+08) (65536,7.4952e+08) (262144,8.09423e+08) (1048576,8.81402e+08) (4194304,4.0806e+08) (16777216,3.05256e+08) };
  \addlegendentry{Fixed};
  \addplot coordinates { (253,5.13821e+08) (511,5.71546e+08) (1021,6.37018e+08) (2047,6.71367e+08) (4093,7.16191e+08) (8191,7.36448e+08) (16381,6.85403e+08) (32767,6.56547e+08) (65533,6.71765e+08) (131071,7.41536e+08) (262141,7.60651e+08) (524287,7.86349e+08) (1048573,8.22131e+08) (2097151,8.45397e+08) (4194301,4.13987e+08) (8388607,3.34057e+08) (16777213,3.14067e+08) };
  \addlegendentry{Variable};

  %% MULTIPLOT(method|title) SELECT "Skewed" || method as title, paths AS x,
  %% edges / avg(total) * 1000 AS y, MULTIPLOT
  %% FROM rmatskewed WHERE type="sampling" AND method != "graph500" AND method != "networkit"
  %% GROUP BY MULTIPLOT, paths ORDER BY MULTIPLOT, paths
  \addplot coordinates { (256,5.77516e+08) (1024,7.04348e+08) (4096,7.91256e+08) (16384,7.42919e+08) (65536,7.49906e+08) (262144,8.09621e+08) (1048576,8.82005e+08) (4194304,4.10213e+08) (16777216,3.06101e+08) };
  \addlegendentry{Fix skewed};
  \addplot coordinates { (253,8.47503e+08) (511,9.94072e+08) (1021,1.11405e+09) (2047,1.14691e+09) (4093,1.17935e+09) (8191,1.21886e+09) (16381,1.12286e+09) (32767,1.03331e+09) (65533,9.9726e+08) (131071,1.05675e+09) (262141,1.05132e+09) (524287,1.04912e+09) (1048573,1.05167e+09) (2097151,1.04479e+09) (4194301,4.71201e+08) (8388607,3.68647e+08) (16777213,3.36237e+08) };
  \addlegendentry{Var skewed};

  %% MULTIPLOT(method|title) SELECT method as title, paths AS x,
  %% edges / avg(total) * 1000 AS y, MULTIPLOT
  %% FROM rmat WHERE type="sampling" AND method="graph500" OR method="networkit"
  %% GROUP BY MULTIPLOT, paths ORDER BY MULTIPLOT, paths
  %\addplot coordinates { (0,6.20978e+07) };
  \addplot+[mark=none, dashed, black] coordinates {(1, 6.20978e+07) (1e8, 6.20978e+07)};
  \addlegendentry{Graph 500};
  %\addplot coordinates { (0,1.11478e+08) };
  \addplot+[mark=none, dotted, line cap=round, black] coordinates {(1, 1.11478e+08) (1e8, 1.11478e+08)};
  \addlegendentry{NetworKit};

  %% PLOT SELECT paths AS x, edges / avg(total) * 1000 AS y
  %% FROM rmatskewed WHERE type="sampling" AND method="networkit"
  %% GROUP BY paths ORDER BY paths
  %\addplot coordinates { (0,1.9932e+08) };
  %% \addplot+[mark=none, dashdotdotted, line cap=round, black] coordinates {(1, 1.9932e+08) (1e8, 1.9932e+08)};
  %% \addlegendentry{NetworKit skewed};
  \end{semilogxaxis}
\end{tikzpicture}
\caption{Throughput of R-MAT generator as a function of table size, using
  64 threads.}
  %Parameters: $n=2^{30}$, $m=10^8$, averaged over 20 iterations with
  %50 repetitions.
\label{fig:tablesize}
\end{figure}
% RANGE END plot

% RANGE BEGIN speedup
% CONNECT sqlite:/global_data/lorenz/wrs/results.db
% SQL CREATE INDEX IF NOT EXISTS rmatscalingtype ON rmatscaling (type)
\begin{figure}
\centering
\begin{tikzpicture}
  \begin{semilogxaxis}[
    log basis x=2,
    log ticks with fixed point,
    %ymode=log,
    xlabel={Threads},
    ylabel={Speedup},
    % legend pos=south east,
    legend pos=north west,
    width=10cm, % for now
    height=6.5cm,
    ymin=-3,
    ymax=34.5
  ]
  %% MULTIPLOT(method|title) SELECT threads AS x, (
  %% SELECT avg(total) FROM rmatscaling WHERE method="fixdepth" AND type="sampling"
  %% AND threads=1 AND scramble=0) / avg(total) AS y, MULTIPLOT,
  %% method as title FROM rmatscaling WHERE scramble=0 AND type="sampling"
  %% GROUP BY MULTIPLOT, threads ORDER BY MULTIPLOT,x
  \addplot coordinates { (1,1.0) (2,1.96871) (4,3.83738) (8,7.15895) (16,15.3348) (32,30.6399) (64,32.0116) };
  \addlegendentry{Fixed};;
  \addplot coordinates { (1,0.810278) (2,1.60441) (4,3.15024) (8,6.35341) (16,12.2621) (32,25.1155) (64,30.6768) };
  \addlegendentry{Variable};
  \addplot coordinates { (1,0.0485554) (2,0.0970334) (4,0.193872) (8,0.38754) (16,0.771013) (32,1.54169) (64,2.25436) };
  \addlegendentry{Graph 500};
  \addplot coordinates { (1,0.0889881) (2,0.177799) (4,0.354944) (8,0.708359) (16,1.40792) (32,2.8049) (64,4.03663) };
  \addlegendentry{NetworKit}

  \addplot+[mark=none, dashed, gray] coordinates {(32, -3) (32, 35)} node[right, pos=0.5] {HT};
  \end{semilogxaxis}
\end{tikzpicture}
\caption{Speedup relative to the fastest sequential algorithm, our fixed depth method.}
\label{fig:speedup}
\end{figure}
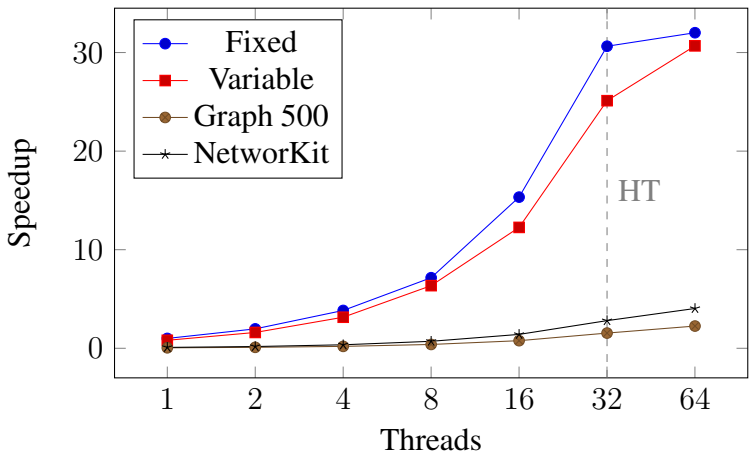
% RANGE END speedup

% RANGE BEGIN stats
% CONNECT sqlite:/global_data/lorenz/wrs/results.db
% SQL CREATE INDEX IF NOT EXISTS rmatscalingtype ON rmatscaling (type)

%% DEFMACRO REFORMAT(precision=1) SELECT
%% (SELECT avg(total) as seqg500 from rmatscaling WHERE method="graph500"
%% AND threads=1 AND type="sampling" AND scramble=0) / avg(total) as fixfactor
%% FROM rmatscaling WHERE method="fixdepth" AND threads=1 AND type="sampling" AND scramble=0
\def\fixfactor{20.6}

%% DEFMACRO REFORMAT(precision=1) SELECT
%% (SELECT avg(total) as seqg500 from rmatscaling WHERE method="graph500"
%% AND threads=64 AND type="sampling" AND scramble=0) / avg(total) as fixfactorall
%% FROM rmatscaling WHERE method="fixdepth" AND threads=64 AND type="sampling" AND scramble=0
\def\fixfactorall{14.2}

%% DEFMACRO REFORMAT(precision=1) SELECT
%% (SELECT avg(total) as seqg500 from rmatscaling WHERE method="networkit"
%% AND threads=64 AND type="sampling" AND scramble=0) / avg(total) as fixfactorallnk
%% FROM rmatscaling WHERE method="fixdepth" AND threads=64 AND type="sampling" AND scramble=0
\def\fixfactorallnk{7.9}

%% DEFMACRO REFORMAT(precision=1) SELECT
%% (SELECT avg(total) as seqg500 from rmatscaling WHERE method="graph500"
%% AND threads=1 AND type="sampling" AND scramble=0) / avg(total) as varfactor
%% FROM rmatscaling WHERE method="vardepth" AND threads=1 AND type="sampling" AND scramble=0
\def\varfactor{16.7}

%% DEFMACRO REFORMAT(precision=1) SELECT
%% (SELECT avg(total) as seqg500 from rmatscaling WHERE method="graph500"
%% AND threads=1 AND type="sampling" AND scramble=1) / avg(total) as fixsfactor
%% FROM rmatscaling WHERE method="fixdepth" AND threads=1 AND type="sampling" AND scramble=1
\def\fixsfactor{10.3}

%% DEFMACRO REFORMAT(precision=1) SELECT
%% (SELECT avg(total) as seqg500 from rmatscaling WHERE method="graph500"
%% AND threads=1 AND type="sampling" AND scramble=1) / avg(total) as varsfactor
%% FROM rmatscaling WHERE method="vardepth" AND threads=1 AND type="sampling" AND scramble=1
\def\varsfactor{9.2}

% RANGE END stats
We perform experiments on a machine with two Intel Xeon E5-2683 v4 processors
which have 16 cores each.  With hyperthreading this means we can use
up to 64 threads.  Our implementation is written in C++ and compiled with the
GNU C++ compiler \texttt{g++} in version 8.2.0.
The source code is available at \url{https://github.com/lorenzhs/wrs/tree/rmat}.
We
compare with the R-MAT generators used in the Graph 500 benchmark reference implementation and in NetworKit \cite{networkit}.
We removed the code for making the graph undirected and for scrambling vertex IDs
in order to concentrate on the core task of R-MAT: edge generation.
The modified code is also available under the above link.

Figure~\ref{fig:tablesize} shows the throughput
using 64 threads as a function of the available table size.
%% In this implementation we control the table size exactly by keeping
%% a max-priority queues of possible table entries. In each iteration,
%% the table entry with maximum probability $p$ is expanded into four
%% new entries with probabilites $\set{ap,bp,cp,dp}$.
For $n=2^{30}$ nodes, we generate $10^{11}$ edges overall,
assigning blocks of $2^{16}$ edges at a time to the threads.
We see two peaks in throughput which correspond to fully using L2 cache and L3 cache respectively.
Overall, on the Intel machine, the best throughput is up to 881
million edges per second.%
\footnote{On an AMD Epyc 7551P with 32 cores we get similar results -- slightly better overall performance, best performance when the L3 cache on each chiplet is fully used.} This is $\fixfactorall$ times faster than the Graph 500 generator and $\fixfactorallnk$ times faster than NetworKit.
Adding the ``clip-and-flip'' technique to produce undirected graphs adds a constant cost of 5--8\,ns per edge\fragelater{oddly, 5 for variable depth and the existing ones, 8 for fixed depth}. Scrambling (to prevent node IDs from giving away information about their likely properties) adds another 40\,ns per edge using our optimised implementation, which is faster than in Graph 500 \fragelater{oddly, for the Graph 500 generator, it's only around 10ns???}.
We can see that the variable length variant performs slightly worse than the fixed length variant since there is some overhead for processing variable length bit strings.

For more skewed parameters, the variable length variant comes out
ahead.  Setting $a=0.9$, $b=c=0.025$ and $d=0.05$, the variable length variant achieves $1.23\cdot10^9$ edges per
second with a table size of $2^{13}$ entries.  This is roughly 40\,\% more than
the fixed depth variant.  Note that for this choice of parameters, utilising the
full L3 cache is no longer optimal: as the entropy of the alphabet is now much
lower, the $2^{13}$ entries that fit into each core's L2 cache provide 13.9 bits
of information on average.  Increasing the table size to $2^{21}$ yields 15 bits
per sample, which is not enough of an improvement to offset the increased
latency of the L3 cache.  Both of these figures compare to exactly 10 bits per
sample for the best-performing fixed depth configuration, whose running time
remains unchanged (it does not depend on the parameters $a$--$d$).

Figure~\ref{fig:speedup} gives the speedup over our algorithm on a single thread.
We achieve speedup 30.6 on 32 threads and speedup 32 on 64 threads.
Hyperthreading is of little help.  This is because the memory bandwidth to L3 cache is becoming a limiting factor in our code. In contrast, the Graph 500 generator significantly benefits from hyperthreading since the number of arithmetic operations far dominate the memory accesses.

The speed of our generators compares favorably to generators for other
models. For example, the sequential Erd\H{o}s-Renyi generator of \cite{FLSSSL18}
is only around 10\,\% faster than our sequential generator, but for a much simpler
model (the special case $a=b=c=d$).
% 9s for 2^28 edges sequentially = 33.5ns per edge on i10pc130, we have ~35 on
% i10pc132, around 5% but let's add another 5% to account for the newer machine
On the other hand, the streaming hyperbolic graph generator sRHG of
\cite{FLSSSL18}, a more complicated model, takes around 5 times more time per
edge. % 180ns per edge (74s for 16e9 edges on 39 cores)

%%%%%%%%%%%%%%%%%%%%%%%%%%%%%%%%%%%%%%%%%%%%%%%%%%%%%%%%%%%%%%%%%%%%%%
\section{Conclusions}
We give a simple and practical algorithm for generating R-MAT graphs with
constant time per edge in an embarrassingly parallel way. This makes this widely
used family of graphs even more easily usable for experiments with huge
graphs. We improve sequential throughput by an order of magnitude over the basic
algorithm by precomputing chains consisting of many random decisions and storing
them in an appropriate discrete distribution. The same approach is likely to
work in other settings too.

\bibliographystyle{elsarticle-num}
\bibliography{diss}

\end{document}